\documentclass[%
superscriptaddress,
preprint,
 amsmath,amssymb,
 aps,
prb,
]{revtex4-2}

\usepackage{mathrsfs}

\usepackage{graphicx}% Include figure files
\usepackage{dcolumn}% Align table columns on decimal point
\usepackage{bm}% bold math
\begin{document}

\title{Mesoscopic Helices of Polar Domains in a Quadruple Perovskite}

%\title{Revisiting Helical Dipole Wave in a Quadruple Perovskite}

%\title{Mesoscopic Helical Ordering of Polar Domains in Copper Doped BiMn$_{7}$O$_{12}$}

\author{Yang Zhang}
\email{yzhang6@fas.harvard.edu}
\affiliation{
The Rowland Institute at Harvard, Harvard University, Cambridge, MA 02138, USA
}%

\author{Mingyu Xu}
\affiliation{
Department of Chemistry, Michigan State University, East Lansing, MI 48824, USA
}%

\author{Jie Li}
\affiliation{
Department of Earth and Environmental Sciences, University of Michigan Ann Arbor, MI 48109, USA
}%

\author{Suk Hyun Sung}
\affiliation{
 The Rowland Institute at Harvard, Harvard University, Cambridge, MA 02138, USA
}%

\author{Sang-Wook Cheong}
\affiliation{
Keck Center for Quantum Magnetism and Department of Physics and Astronomy, Rutgers University, New Jersey 08854, USA
}%

\author{Weiwei Xie}
\affiliation{
Department of Chemistry, Michigan State University, East Lansing, MI 48824, USA
}%

\author{Ismail El Baggari}
\email{ielbaggari@fas.harvard.edu}
\affiliation{
 The Rowland Institute at Harvard, Harvard University, Cambridge, MA 02138, USA
}%

\date{\today}% It is always \today, today,

\begin{abstract}
A significant effort in condensed matter physics is dedicated to the search for exotic arrangements of electric dipoles in crystals. 
Non-collinear dipolar arrangements mimicking magnetic spin textures, such as polar vortices and skyrmions, have been realized, but others, like helices of dipoles, have remained elusive.
%Contrary to a previous report, helical order in this material has not yet been shown, motivating a renewed search for helical textures. 
While earlier work claimed the presence of a helical dipole modulation in ferroelectric BiCu$_{x}$Mn$_{7-x}$O$_{12}$, our results rule out such a structure, motivating a renewed search for helical textures.
Using atomic-resolution imaging, we report a novel form of helical order in which polar domains self-organize into a mesoscopic helical pattern:
from domain to domain, polarization rotation follows a consistent handedness.
Both right- and left-handed chirality emerging from this helical ordering are observed.
This discovery establishes mesoscopic ordering as novel mechanism for inducing helical textures, and hence chirality, in ferroelectric crystals, paralleling the efforts in liquid crystals and supramolecular assemblies.
%., with a broad applications optical activity in ferroelectric crystals.
%Our work also serves as a broader caution that the dense domain structures in materials can mimic complex structural motifs if not corroborated with high spatial techniques resolution.
\end{abstract}

\maketitle

\newpage
\subsection*{Introduction}
Spin arrangements in magnets underlie diverse functional properties. 
The simplest configurations have aligned (ferromagnet) or anti-aligned (antiferromagnet) spins, but these can also break into complex domain patterns \cite{shull1949detection,shpyrko2007direct}.
Magnetic systems are a fertile ground for realizing much more exotic non-collinear spin textures, such as cycloids, helicoids, or skyrmions \cite{yu2010real,zheng2023hopfion}. 
Such exotic spin spirals have emerged as an integral part to realizing multiferroic functionality that couples electrical, magnetic, and structural degrees of freedom.
For example, cycloidal spirals of spins are found in TbMnO$_{3}$, Ni$_{3}$V$_{2}$O$_{6}$, and MnWO$_{4}$\cite{kimura2003magnetic,kenzelmann2005magnetic,katsura2005spin,mostovoy2006ferroelectricity,spaldin2019advances,masuda2021electric}.
%For example, cycloidal spirals of spins, $\mathbf{S_i}$, can lead to polarization, $\mathbf{P}$, given by $\mathbf{P} \sim \mathbf{r_{ij}} \times [\mathbf{S_i} \times \mathbf{S_j}] $ and where $\mathbf{r_{ij}}$ is the vector connecting $\mathbf{S_i}$ and $\mathbf{S_j}$.
%Canonical examples of spin spirals include TbMnO$_{3}$, Ni$_{3}$V$_{2}$O$_{6}$, and MnWO$_{4}$

Electrical dipoles also form aligned (ferroelectric) or anti-aligned order (antiferroelectric), as well as complex domain configurations \cite{sawaguchi1951antiferroelectric,merz1954domain,van2007observation,jia2011direct}.  
While it is comparatively difficult to realize exotic electric dipole textures, advances in heterostructure engineering have created topological structures like polar vortices and merons \cite{yadav2016observation,das2019observation,wang2020polar,shao2023emergent,hu2024giant}. 
But unlike in magnetic systems, the realization of chiral polar helices in ferroelectric crystals has been far less successful and remains a hotly pursued effort.

%These textures are achieved by carefully balancing the competition between strain, electrostatics, and polar instabilities \cite{hong2017stability,junquera2023topological}. 
%Despite this, the complexity of fine tuning these interactions motivates a parallel effort to discover bulk materials that can intrinsically host complex textures of electric dipoles, which could in turn inform new design principles for intricate ferroic states.

To date, helical dipolar textures have been primarily discovered in liquid crystals and supra-molecular assemblies, remaining a vibrant research topic for many years \cite{bahr2001chirality, yashima2016supramolecular}.  
Discoveries and observations of chiral helical ground states in these systems are not only of fundamental interest but also hold potential for novel electro-optical applications \cite{tortora2011chiral,nayani2015spontaneous,chen2020first,kumari2024chiral, karcz2024spontaneous}.
In crystalline ferroelectrics, however, chiral helices of electric dipoles textures have not been observed.
A notable exception is the report of a possible helical polar texture in a so-called quadruple perovskite BiCu$_{0.1}$Mn$_{6.9}$O$_{12}$ in the form of periodic twisted atomic displacements \cite{khalyavin2020emergent}.

Using advanced atomic-resolution electron microscopy, we show that the search of polar helices must continue, as the previously resolved helical structure is an artifact of structural refinements.
Superlattice diffraction spots, purported to reflect an incommensurate dipole helical wave in BiCu$_{0.1}$Mn$_{6.9}$O$_{12}$, are instead a reflection of a high density of polar structural domains visible only under high-resolution imaging.
Remarkably, we uncover instead a novel form of long-range helical order - not in the atomic displacements themselves, but in the mesoscale arrangement
of these domains. 
Crossing from one domain to the next, the direction of
polar displacement rotates clockwise (or counterclockwise) around the propagation direction, creating
a mesoscopic helical pattern of polar rotation. 
This behavior reveals a striking emergent
self-organization on the mesoscale and provides intriguing opportunities for realizing novel chiral helical textures from otherwise non-chiral building blocks in ferroelectric crystals.
% To the best of our knowledge, this is the first work that has claimed to observe helical polar textures in a crystalline ferroelectric. 
%Such helical dipole modulation is so promising that it is viewed as a realization of structural chirality in ferroelectric crystals, mimicking those of liquid crystals and supramolecular structures \cite{huck1996dynamic,link1997spontaneous, kumari2024chiral, karcz2024spontaneous}.

\subsection*{Results}
BiCu$_{0.1}$Mn$_{6.9}$O$_{12}$ is a so-called quadruple perovskite lightly doped with copper.
In the parent BiMn$_{7}$O$_{12}$, Bi is 12-fold-coordinated and Mn occupies two distinct coordination sites: 
one in square-planar-coordination and another in octahedral coordination similar to perovskites (Fig. 1A) \cite{maia2024two}.
Below $\sim$ 390 K, BiMn$_{7}$O$_{12}$ hosts two low-symmetry polar phase transitions, a $Cm$ phase below $\sim$ 390 K and a $P1$ phase at room temperature, both promoting polar displacement of Bi due to the electronic instabilities of the 6s$^{2}$ lone pair of Bi$^{3+}$ \cite{slawinski2017triclinic,belik2017complex}.
The polar displacements of the Bi cations aligns close to the [$1\bar{1}0$]$_{pc}$ direction in $Cm$ phase (right panel of Fig. 1A, all crystallographic axes in the polar phase are based on the pseudocubic system), while it shows a subtle rotation around [001]$_{pc}$ axis in $P1$ phase \cite{soboleva2024understanding,maia2024two}. 

After light copper doping, two novel low-temperature phase transitions that radically differ from those in BiMn$_{7}$O$_{12}$ have been proposed, based on X-ray and neutron studies \cite{khalyavin2020emergent}.
Specifically, a helical dipole wave in ferroelectric crystals was claimed to be discovered, characterized by strong incommensurate satellite reflections in reciprocal space.
The proposed helical dipole wave features continuous rotation of Bi atomic displacement in a helical pattern along an axial vector along the pseudocubic $[111]$$_{pc}$ direction (Fig. 1B).

We first confirm the incorporation and concentration of copper in our sample through X-ray diffraction and energy dispersive X-ray spectroscopy (see Tables S1-S4 and Fig. S1).
Figure 1C shows a simulated high-angle annular dark field STEM (HAADF-STEM) images of BiCu$_{0.1}$Mn$_{6.9}$O$_{12}$ viewed along [$11\bar{2}$]$_{pc}$ zone axis.
In the HAADF-STEM imaging mode, electrons that have scattered to high angles are collected and so the image intensity scales monotonically with the atomic number.
Bi atomic columns appear brighter than Mn, while O columns are not visible due to their low scattering cross section.  
From to the refined atomic positions, the proposed helical atomic modulations of bismuth along the [111]$_{pc}$ direction are large, reaching up to 0.4 \r{A} in amplitude.
Therefore, a striking modulation on Bi positions should be visible in HAADF-STEM image when viewed along the [$11\bar{2}$]$_{pc}$ axis (left panel of Fig. 1C).
However, our experimental atomic-resolution data (Fig. 1D) show no such modulated bismuth positions; 
instead, the structure matches that of the undoped parent material (right panel of Fig. 1C), thus revealing an unexpected divergence from the refined atomic structure.

Stark differences between the undoped and Cu-doped samples, however, become evident at mesoscopic length scales. 
Figures 1E and F display $250 \times 250~\mathrm{nm}^2$ low-angle ADF STEM (LAADF-STEM) images of both samples along [$1\bar{1}0$] axis. 
The LAADF scattering signal includes electrons that have scattered to low angles and is more sensitive to lattice strain and structural defects, leading to enhanced contrast 
\cite{muller2004atomic, lee2018isostructural}. 
As shown in Figs. 1E and F, a high density of bright lines appears across the field-of view in BiCu$_{0.1}$Mn$_{6.9}$O$_{12}$, a feature consistently observed in multiple samples (Fig. S2). 
In comparison, the undoped sample exhibits either a complete absence of such line contrasts or a much lower density over similar fields of view (Fig. S3). 
The lines are oriented perpendicular to the [$\bar{1}\bar{1}1$]$_{pc}$ direction, matching the direction of incommensurate superlattice peaks (inset of Figs. 1E and F, raw data shown in Fig. S4).
A detailed analysis of the superlattice peaks reveals that the incommensurate peaks, previously attributed to a helical displacive modulation of Bi atoms, relate instead to the domain boundaries.
Regions lacking domains show no incommensurate peaks, whereas regions with line features consistently exhibit them (right panel of Fig. 1F).
Quantitative analysis based on 4D-STEM (see Methods for details) reveals that all of these line contrasts correspond to boundaries between ferroelastic domains, across which the lattice spacings along [001]$_{pc}$ and [110]$_{pc}$, as well as the crystallographic angles, undergo distinct changes (Fig. 1G and Fig. S5).

Typically, an incommensurate superlattice peak in diffraction indicates the presence of a real-space modulation whose periodicity is not an integer multiple of the underlying unit cell.
Three-dimensional translational periodicity is broken, and structural refinements must use higher-dimensional superspace groups.
However, the solution space for incommensurate structures is highly degenerate, making structure determination particularly challenging.
This complexity is exacerbated by the presence of real-space features within the crystal.
Notably, incommensurate superlattice peaks may reflect localized features in a crystal, such as discommensurations seen in charge order systems \cite{mesaros2016commensurate,el2018nature,schnitzer2025atomic}.
In our case, sharp domain boundaries in the crystal create intense superlattice reflections that are incommensurate with the lattice periodicity. 
Summarizing thus far, our results indicate that BiCu$_{0.1}$Mn$_{6.9}$O$_{12}$ preserves the underlying crystallographic structure of BiCuMn$_{7}$O$_{12}$, and that the primary effect of copper doping is the substantial increase in the density of structural domains.

As these domains are ferroelastic, we next explore their interplay with polar Bi displacements.
In the pseudocubic cell, there are three (or six if we distinguish between positive and negative directions) inequivalent $<$110$>$$_{pc}$ orientations lying on the ($\bar{1}\bar{1}1$)$_{pc}$ pseudocubic plane (labeled I, II and III in Fig. 2A).
By precisely fitting and mapping atomic positions using HAADF-STEM along [$1\bar{1}0$]$_{pc}$ axis (Fig. 2B), we extract (i) local bond distances, \textit{$d_{[110]pc}$} and (ii) polar displacements of Bi atoms relative to the Mn sublattice, $\mathbf{\Delta}_{\mathrm{Bi}}$, as defined in the right panel of Fig. 2B.
Figure 2C shows differences in \textit{$d_{[110]pc}$} when polar displacements are along three inequivalent $<$110$>$$_{pc}$ orientations, revealing ferroelastic domains.
Type-I domain is evident due to its larger \textit{$d_{[110]pc}$} of 5.47 \text{\AA}. 
Type-II and III domains have a smaller \textit{$d_{[110]pc}$}, but we cannot be reliably distinguish between them since in projection they have similar lattice constants and are close to the precision limit of atomic fitting.

Unlike the lattice constants, polar displacements of Bi have characteristic signatures that uniquely determine all ferroelastic domain variants.
Although STEM is a projection imaging technique, the chosen viewing direction identifies a distinct in-plane component of the polar displacement for each of the domains. 
As shown in the atomic model (Fig. 2C) and experimental measurement (Fig. 2D), Type-II and Type-III domains exhibit a significant yet distinct in-plane components of $\mathbf{\Delta}_{\mathrm{Bi}}$, identifiable through finite amplitude and distinct directions. 
$\mathbf{\Delta}_{\mathrm{Bi}}$ in Type-I domains is largely along the imaging direction, yet it still has a measurable, albeit smaller, in-plane component.
Figure 2E shows polar histograms of $\mathbf{\Delta}_{\mathrm{Bi}}$, revealing that each of the three types of polar domains has a characteristic amplitude and angle of the in-plane projection of the displacement. 
Detailed multislice simulations (see Methods and Fig. S6) provide additional confirmation of the accurate identification of the three polar domains.

Figure 2F shows a map of \textit{$d_{[110]pc}$} from $50 \times 35~\mathrm{nm}^2$ region, which spans multiple domains (also see Fig. S7).
Figure 2G shows a $\mathbf{\Delta}_{\mathrm{Bi}}$ map from the same region, with the color and transparency reflecting the direction and amplitude, respectively.
While only two domains are visible in the \textit{$d_{[110]pc}$} map, the $\mathbf{\Delta}_{\mathrm{Bi}}$ unearths all three domains.
Further inspection of Fig. 2G shows a remarkable result. 
The sequence of domains follows a repeating  "III-II-I" pattern along [$\bar{1}\bar{1}1$]$_{pc}$ direction. 
In three dimensions, this corresponds to a helical self-organization of polar domains over mesoscopic length scales (Fig. 2H). 
Across each domain boundary along [$\bar{1}\bar{1}1$$_{pc}$ direction, the polar displacement undergoes a 120-degree counterclockwise rotation within the ($\bar{1}\bar{1}1$)$_{pc}$ plane. 
Here, the sense of rotation is defined when viewed along [$\bar{1}\bar{1}\bar{1}$]$_{pc}$.
This produces an asymmetry within the Type-I domains: on one side, they are always preceded by a Type-II domain and, on the other, followed by a Type-III domain.
These arrangements extend over hundreds of nanometers and are consistently observed across multiple samples and regions (Fig. S8), indicating a long-range mesoscale helical ordering of polar domains.
Such a helical state is an unusual realization of structural chirality, manifesting in this case as a left-handed variant along the [$\bar{1}\bar{1}1$]$_{pc}$ direction.
This mesoscopic helix of polar domains is chiral because it is not superimposable on its mirror image.

An obvious question to ask is whether such polar domains can form a helical ordering with the opposite chirality, as shown in Fig. 3A.
Upon conducting additional searches, we discovered the presence of opposite right-handed chirality in the polar domain ordering. 
Figure 3B shows a 4D-STEM map of polarization (see Methods and Figs. S9-S11), further processed with difference of Gaussians to better visualize the three domains and their consistent helical arrangement over large scales. 
Long-range helical orderings with opposite chirality are observed: 
the color sequence in the right panel is reversed when compared to the left-handed counterpart shown in the left panel. 
Further atomic-scale mapping of a sub-region (Fig. 3C) reveals helical order which involves a clockwise rotation of polar order across domains.
This clockwise (right-handed chiral) ordering comprises polar displacements that are 180° reversed relative to the counterclockwise (left-handed chiral) ordering, as seen in the polar histogram in Fig. 3D.
These real-space observations thus establish that both chiral states can emerge from mesoscopic helices of polar domains.

\subsection*{Discussion}
In ferroelectric materials, electrostatic and elastic energy considerations drive the formation of polar domain configurations that minimize the system's total energy. 
In this study, we report the discovery of a novel form of helical dipole wave in a bulk crystal, not characterized by a continuous rotation at atomic scale, but by a long-range arrangement of polar domains. 
This mesoscale helical arrangement imparts chirality to the material, with both left- and right-handed variants observed, mimicking the chiral ground states observed in liquid crystals \cite{kumari2024chiral, karcz2024spontaneous}.
Mesoscale features—such as domains and domain walls—play a key role in enhancing materials' functionalities \cite{seidel2009conduction, mundy2017functional, farokhipoor2014artificial, ma2018controllable}. 
Therefore, the emergence of helical domain configurations in BiCu$_{x}$Mn$_{7-x}$O$_{12}$ provides a new opportunity for engineering chiral textures through mesoscopic ordering of polar building blocks.  

A second remarkable finding from this atomic-resolution study is how light copper doping dramatically tunes the density of mesoscopic domains.
While the precise mechanism by which copper modulates the domain density remains an open question, the dramatic effect herein presents an opportunity to tune the density and period of mesoscopic textures. 
A comprehensive doping study has shown that copper doping levels up to $x$=1.2 are linked  to a multitude of macroscopic properties including a re-entrant high-temperature phase with near-zero thermal expansion \cite{belik2017reentrant, belik2021plethora}. 
If our visualizations are any indication, the rich phase diagram of BiCu$_{x}$Mn$_{7-x}$O$_{12}$ could reflect a range of mesoscopic domain configurations, suggesting a potential to tailor macroscopic responses through domain engineering of bulk crystals.
Finally, these results highlight the power of atomic-resolution imaging in uncovering the precise structure of ordered phases.

\section*{Acknowledgments}
We acknowledge discussions with Yu-Tsun Shao, Chuqiao Shi, Xinyan Li, Robert Hovden, Rose Hosking, Andrew W. Murry, Michele Conroy, John Heron, Pu Yu, Houbing Huang and Bo Wang. 
\textbf{Funding:} Y. Z., S. H. S and I. E were supported by the Rowland Institute at Harvard.
The high-pressure synthesis and characterization at MSU and University of Michigan were supported by NSF-DMR-2422361 and NSF-DMR-2422362, respectively.
Focused ion beam sample preparation was performed at the Harvard University Center for Nanoscale Systems (CNS); a member of the National Nanotechnology Coordinated Infrastructure Network (NNCI), which is supported by the National Science Foundation under NSF award no. ECCS-2025158. 
S.-W. C. was supported by the W. M. Keck foundation grant to the Keck Center for Quantum Magnetism at Rutgers University.
Transmission electron microscopy was carried out through the use of MIT.nano facilities.
\textbf{Author contributions}
Y. Z., I.E. and S.-W. C. conceived the project.
Y. Z. and I.E. performed electron microscopy and data analysis with the help of S. S;
M. X., J. L., S.-W. C. and W. X. synthesized samples;
M. X., J. L. and W. X. performed X-ray diffraction and SEM measurements;
Y. Z. and I. E. wrote the manuscript with input from all authors;
\textbf{Competing interests:} Authors declare no competing interests; 
\textbf{Data and materials availability:} All data are available in the manuscript or the supplementary materials

\bibliography{references}

\newpage

\begin{figure}
    \centering
    \includegraphics[width=\linewidth]{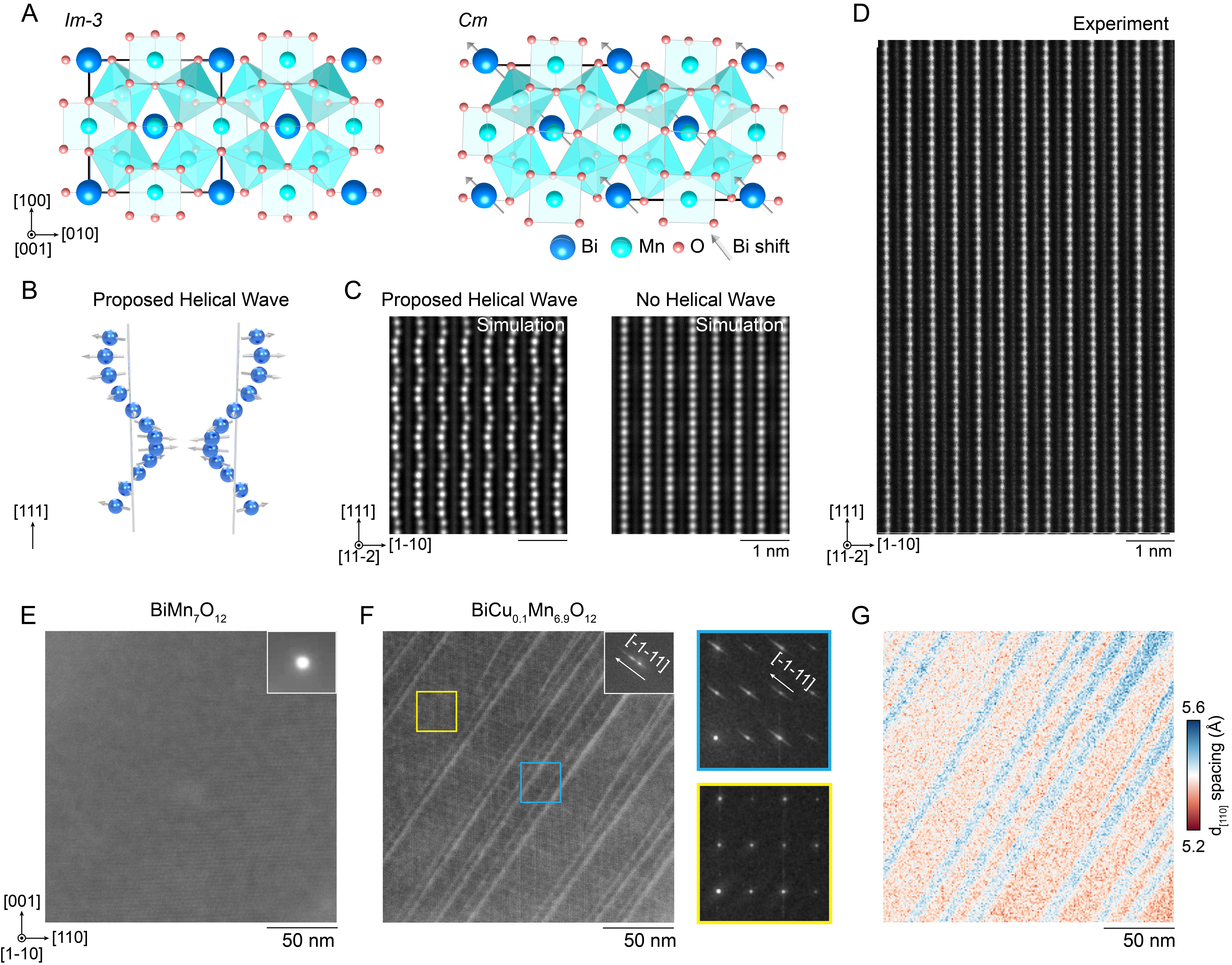}
    \caption{\textbf{Emergence of structural domains with copper doping.}
    (A) Atomic structure of BiMn$_{7}$O$_{12}$ in the $Im\bar{3}$ (left panel) and $Cm$ (right panel) phase.
    The gray arrow in $Cm$ phase shows the polar displacement of Bi along $[1\bar{1}0]$$_{pc}$.
    All crystallography indexing are based on the pseudocubic lattice.
    (B) Previously proposed incommensurate dipole helical wave in BiCu$_{0.1}$Mn$_{6.9}$O$_{12}$.
    Left and right-handed chiralities are shown.
    (C) Simulated ADF-STEM image of structures with (left panel) and without (right panel) dipole helical wave.
    The projection is along [$11\bar{2}$]$_{pc}$
    (D) Experimental ADF-STEM of BiCu$_{0.1}$Mn$_{6.9}$O$_{12}$ along [$11\bar{2}$]$_{pc}$ shows no helical Bi modulation.
    (E, F) $250 \times 250~\mathrm{nm}^2$ field-of-view LAADF-STEM image of (E) BiMn$_{7}$O$_{12}$ and (F) BiCu$_{0.1}$Mn$_{6.9}$O$_{12}$, revealing the presence of high density of domain boundaries in the latter.
    Inserts are the (114)$_{pc}$ Bragg peak. 
    An incommensurate peak along the $[\bar{1}\bar{1}1]$$_{pc}$ direction is evident only in the doped system.
    The right panel of (F) is the Fourier transform from regions highlighted by yellow and blue rectangles.
    The incommensurate peak is absent without a domain boundary (lower panel).
    (G) Lattice spacing along [110]$_{pc}$ in BiCu$_{0.1}$Mn$_{6.9}$O$_{12}$ measured by 4D-STEM (see SI for more details). 
    The field-of view is the same as in (F).
    }
    \label{fig1}
\end{figure}

\begin{figure}
    \centering
    \includegraphics[width=\linewidth]{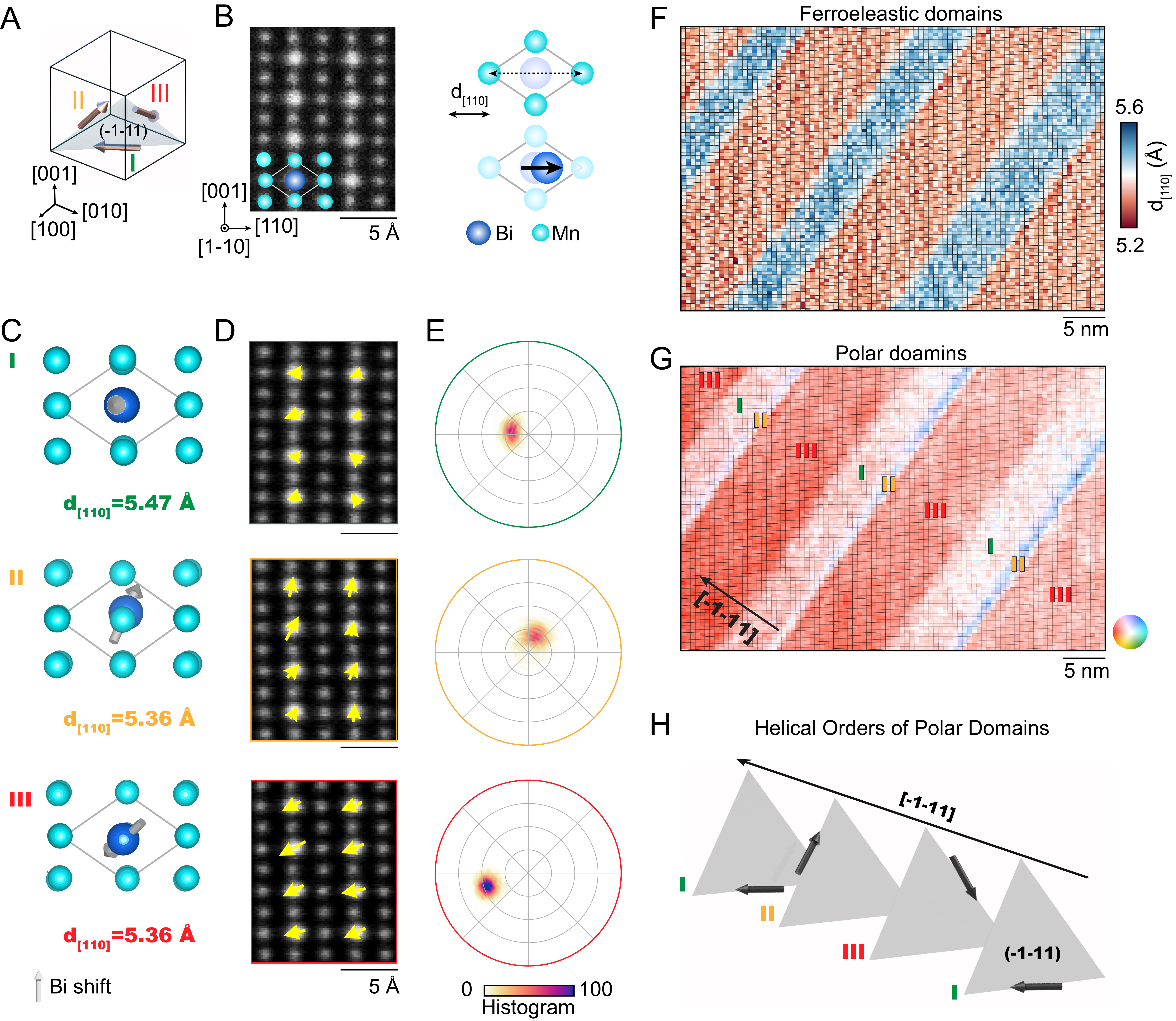}
    \caption{\textbf{A mesoscopic helix of polar domains.}
    (A) Schematic graph of three distinct $<110>_{\text{pc}}$ polar displacements lying on the ($\bar{1}\bar{1}1$)$_{pc}$ plane, corresponding to the three ferroelastic domains (I, II and III).
    (B) HAADF-STEM image of BiCu$_{0.1}$Mn$_{6.9}$O$_{12}$ viewed along [$1\bar{1}0$]$_{pc}$.
    Right panel is the definition of projected lattice spacing along [110] (\textit{d$_{[110]pc}$}) and Bi polar displacement ($\mathbf{\Delta}_{\mathrm{Bi}}$) measured in experiment.
    (C) Atomic structures viewed along [$1\bar{1}0$]$_{pc}$ directions in three ferroelastic domains (I, II and III).
    The gray arrow shows $\mathbf{\Delta}_{\mathrm{Bi}}$.
    The measured \textit{d$_{[110]pc}$} is at the bottom.
    (D) HAADF-STEM image overlaid with measured $\mathbf{\Delta}_{\mathrm{Bi}}$ in three ferroelastic domains.
    (E) Polar histogram of $\mathbf{\Delta}_{\mathrm{Bi}}$ in three ferroelastic domains (I, II and III).
    The radius of the histogram 40 pm.
    Each polar displacement vector has unique planar projection.
    (F) \textit{d$_{[110]pc}$} from $50 \times 35~\mathrm{nm}^2$ field-of view HAADF-STEM image.
    (G) Projected $\mathbf{\Delta}_{\mathrm{Bi}}$ measured from same field-of view HAADF-STEM image as (F).
    The color and transparency represent the polar direction and amplitude, respectively.
    (H) Schematic representation of ordering of polar domains.
    }
    \label{fig2}
\end{figure}

\begin{figure}
    \centering
    \includegraphics[width=\linewidth]{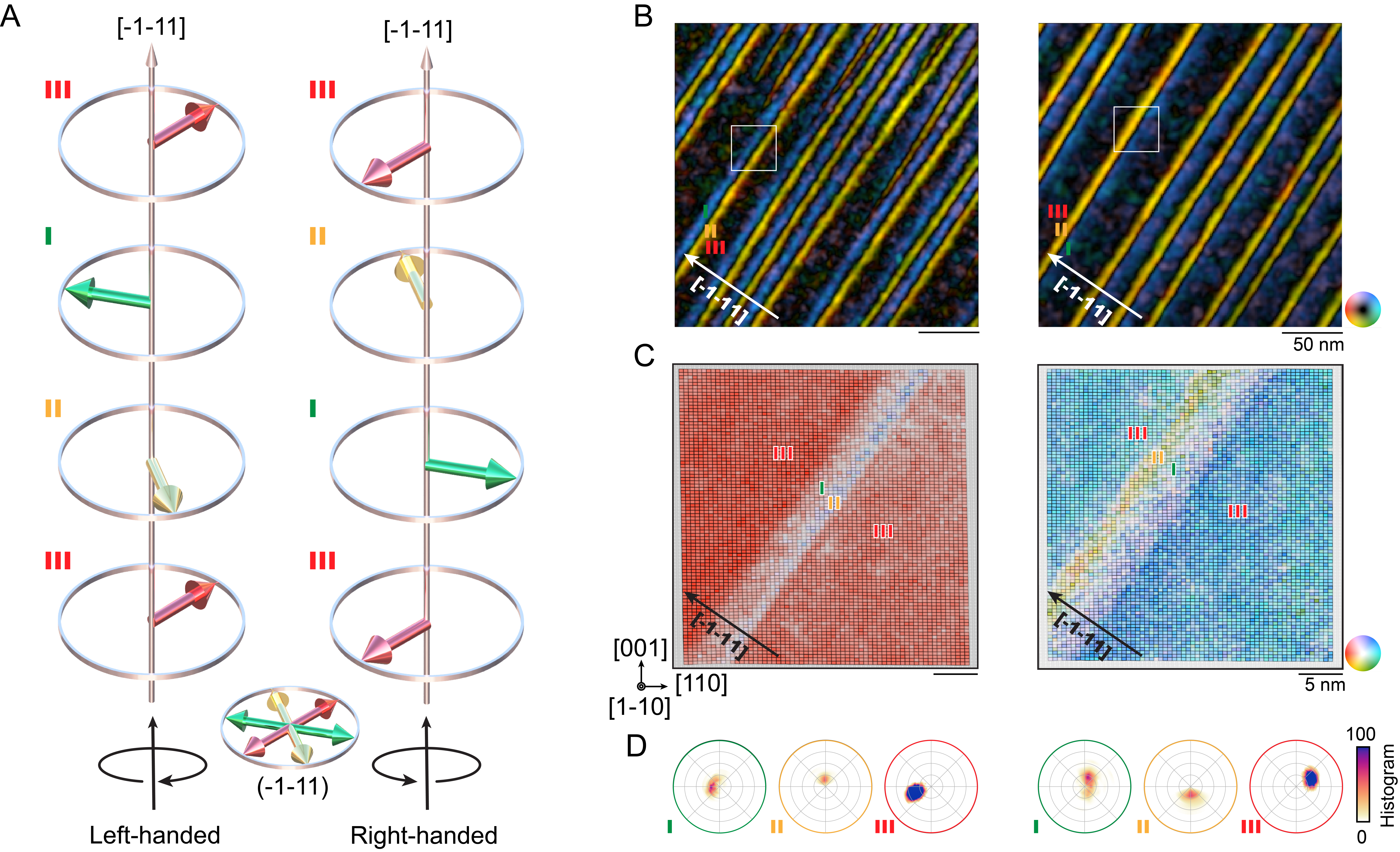}
    \caption{\textbf{Left- and right-handed chirality of mesoscale polar domains}.
    (A) Schematic graph shows left-handed and right-handed chirality, coupled to helical order of polar domains.
    (B) Difference of Gaussian applied to large scale 4D-STEM map of polarization reveals long-range repeat of right- (left panel) and left-handed (right panel) helical ordering of polar domains. 
    Domains I (green), II (yellow) and III (red) are labeled.
    (C) Atomic-scale mapping of polar displacement of sub-regions regions highlighted by white rectangles in (A). 
    (D) Polar histograms of displacements collected from each domain.
    The radius of the histogram is 40 pm.
    }
    \label{fig3}
\end{figure}

% Include the supplementary information

\end{document}